\documentclass[ secnumarabic, amssymb, amsmath,nobibnotes, aps, prl,nofootinbib]{revtex4}

\usepackage{graphicx}
\usepackage{tabularx}

\newcommand{\half}{\mbox{\small{$\frac{1}{2}$}}}
\newcommand{\threehalves}{\mbox{\small{$\frac{3}{2}$}}}
\newcommand{\fourth}{\mbox{\small{$\frac{1}{4}$}}}

\begin{document}
\title{The Khuri-Jones Threshold Factor as an Automorphic Function}
\author{B. H. Lavenda}
\email{info@bernardhlavenda.com}
\homepage{www.bernardhlavenda.com}
\affiliation{Universit$\grave{a}$ degli Studi, Camerino 62032 (MC) Italy}
\begin{abstract}
The Khuri-Jones correction to the partial wave scattering amplitude at threshold is an automorphic function for a dihedron. An expression for the partial wave amplitude is obtained at the pole which the upper half-plane maps on to the interior of  semi-infinite strip. The Lehmann ellipse  exists  below threshold for bound states. As the system goes from below to above threshold, the discrete dihedral (elliptic) group of Type 1 transforms into a Type 3 group, whose loxodromic elements leave the fixed points $0$ and $\infty$ invariant. The transformation of the indifferent fixed points from $-1$ and $+1$ to the source-sink fixed points $0$ and $\infty$ is the result of a finite resonance width in the imaginary component of the angular momentum. The change in symmetry of the groups, and consequently their tessellations, can be used to distinguish bound states from resonances.  
\end{abstract}
\maketitle
\section{Introduction and Summary}
This paper suggests that the origin of strong interaction symmetries may be found in functions that are automorphic with respect to a group generated by a fractional linear transformation. Examples of automorphic functions are the circular functions which are automorphic with respect to $\{2\pi n|n\in\mathbb{Z}\}$, and elliptic functions which are periodic with respect to a group generated by two translations. Elliptic functions live on tilings that are parallelograms which fill up the entire plane. By cutting and pasting, the tessellations can be made into a torus of genus $1$. 

Poincar\'e and Klein were interested in looking for automorphic functions of higher genus; that is,  complex-valued functions invariant under specific groups. Klein had already examples in the shape of regular solids, of which there are only a finite number. We will show that, for real angular momenta, the partial wave amplitude is such an automorphic function corresponding to the dihedral group. According to Khuri~\cite{Khuri} and Jones~\cite{Jones}, the partial wave amplitude derives its form from the asymptotic large angular momentum limit of the Legendre function of the second kind which has three singular points. These singular points are homologues of vertices of triangles in a conformal mapping, and the group we will be dealing with is the triangular group.

In the unphysical region the angular momentum becomes complex, and, consequently, the vertex angles of the triangles in the division of the sphere, which would otherwise have the form of a double pyramid, also become complex.The real part corresponds to a rotation, while the imaginary part represents a stretching. What were indifferent fixed points of an elliptic transformation below threshold, $z=\pm1$, where $z=\cos\vartheta$, become source-sink fixed points, $z=0,\infty$, of a loxodromic transformation above threshold. What mathematically can be obtained by conjugation has a completely different physical explanation: the transition from the physical, $z\in(-1,+1)$, to the unphysical, $z\notin(-1,+1)$, region as the system passes through the threshold. This can also be viewed as a transformation from the discrete, elementary group of Type $1$, consisting of elliptic elements whose group is conjugate to a subgroup of $SO(3)$, to a discrete, elementary group of Type $3$, with loxodromic elements that leave invariant the fixed points $0,\infty$. In this way, it may be feasible to study the strong interaction symmetries through the tessellations of the hyperbolic plane, $\mathbb{H}^2$, and ball, $\mathbb{B}^3$, that Poincar\'e used to study the symmetries of \emph{kleinian\/} groups, or discrete groups of  M\"obius transformations with complex coefficients.

\section{Modified Scattering Amplitude}
It is well-known that the scattering amplitude can be cast into what is now known as the Watson--Sommerfeld representation~\cite{Omnes},
\begin{equation}
A(k,z)=\sum_{i}\frac{\beta_i(k)}{\sin\alpha_i(k)\pi}P_{\alpha_i(k)}(-z), \label{eq:WS}
\end{equation}
although its history goes back much further to Poincar\'e, who used it to study the bending of electromagnetic waves by a sphere~\cite{Poincare}.
The poles occur at integer values of the angular momentum, $\ell=\alpha(k)$, and the residue is written as the product of the Legendre function of the first kind, $P_{\ell}(-z)=(-1)^{\ell}P_{\ell}(z)$,   $\ell=\mbox{integer}$, which contains all the angular dependencies, and a factor $\beta(k)$, which is a function of the linear momentum, $k$. 

Partial waves can be projected out of  \eqref{eq:WS} by using
\begin{equation}
\half\int_{-1}^{1}P_{\ell}(z)P_{\alpha}(-z)dz=\frac{\sin\alpha\pi}{\pi(\alpha-\ell)(\alpha+\ell+1)}, \label{eq:orth}
\end{equation}
for integer $\ell$ and complex $\alpha$. The contribution from one Regge pole to the partial wave scattering amplitude is
\[A_{\ell}(k)=\frac{\beta(k)}{\pi(\alpha(k)-\ell)(\alpha(k)+\ell+1)}.\]
The residue has been found to vary as~\cite{Barut,Newton}
\[\beta(k)\sim k^{2\alpha}.\]
This gives incorrect threshold behavior by predicting that
\[A_{\ell}(0)\sim\frac{k^{2\alpha(0)}}{\pi(\alpha(0)-\ell)(\alpha(0)+\ell+1)}.\]

In order to correct the result, Khuri~\cite{Khuri} and Jones~\cite{Jones} modified the residue, 
\begin{equation}
\beta(k)\rightarrow\beta(k)\exp\left\{-(\ell-\alpha)\cosh^{-1}(z)\right\}, \label{eq:beta}
\end{equation}
where
\begin{equation}
z=\cos\vartheta=1+\frac{m^2}{2k^2}>1,\label{eq:cos}
\end{equation}
in the unphysical region where $k^2>0$, and $m$ is the particle's mass. In the limit as $k^2\rightarrow0$, they obtained the correct threshold behavior  
\[\beta e^{-(\ell-\alpha)\cosh^{-1}(z)}\sim k^{2\alpha}\left(\frac{k}{m}\right)^{2(\ell-\alpha)}=\frac{k^{2\ell}}{m^{2(\ell-\alpha)}}.\]
The added factor \eqref{eq:beta} supplies the partial wave with a branch point at $k^2=-m^2/4$ because the square root in
\[\cosh^{-1}(z)=\ln\left\{1+\frac{m^2}{2k^2}+\surd\left[\left(1+\frac{m^2}{2k^2}\right)^2-1\right]\right\}\]
vanishes there, but it does not give the additional branch points at $-(jm)^2/2k^2$ for $j=2,3,\ldots$.
\section{The Dihedral Group}
Unwittingly, Khuri and Jones transformed the partial wave scattering amplitude into an automorphic function. It is the form that the Legendre function of the second kind takes in the limit as $\ell\rightarrow\infty$~\cite{Erdelyi}, viz.,
\begin{equation}Q_{\ell}(z)\stackrel{|\ell|\rightarrow\infty}{\longrightarrow}\left(1/\surd\ell\right)e^{-(\ell+\half)\cosh^{-1}z}.\label{eq:Q-asy}
\end{equation} Said differently, the automorphic function,
\begin{align}
w&=\left[z+\surd(z^2-1)\right]^{-(\ell-\alpha)}\\
&=\left(\frac{z-\surd(z^2-1)}{z+\surd(z^2-1)}\right)^{\half(\ell-\alpha)}, \label{eq:w}
\end{align}
is the ratio of two independent solutions of the Fuchsian differential equation,
\[
\frac{d^2y}{dz^2}+I(z)y=0,\]
where
\begin{equation}
2I(z)=\{w,z\}=\frac{1-\fourth}{2(1-z)^2}+\frac{1-\fourth}{2(1+z)^2}-\frac{(\ell-\alpha)^2+\half}{2(z^2-1)}.\label{eq:Schwarz}
\end{equation}
The curly brackets denote the Schwarzian derivative of $w$ with respect to $z$,  
\[\{w,z\}=\frac{w^{\prime\prime\prime}}{w^{\prime}}-\threehalves\left(\frac{w^{\prime\prime}}{w^{\prime}}\right)^2.\]
It is a projective invariant that was first discovered by Lagrange in his studies of  conformal mapping of a sphere onto a plane, although the name was coined by Cayley in favor of Schwarz.

 The Schwarzian \eqref{eq:Schwarz} clearly shows that $Q_{\ell}(z)$ has three singular points at $z=-1,+1,\infty$, which are homologues of the vertices of a triangle in the $w$-plane. Since the sum of angles,
\[\left[\half+\half+(\ell-\alpha)\right]\pi>\pi,\]
the triangles are spherical, and there is no \emph{real\/} orthogonal disc which  encloses the triangular tessellations. Hence, the need to project them onto a sphere.

\begin{figure}[ht]
	\centering
		\includegraphics[width=0.4\textwidth]{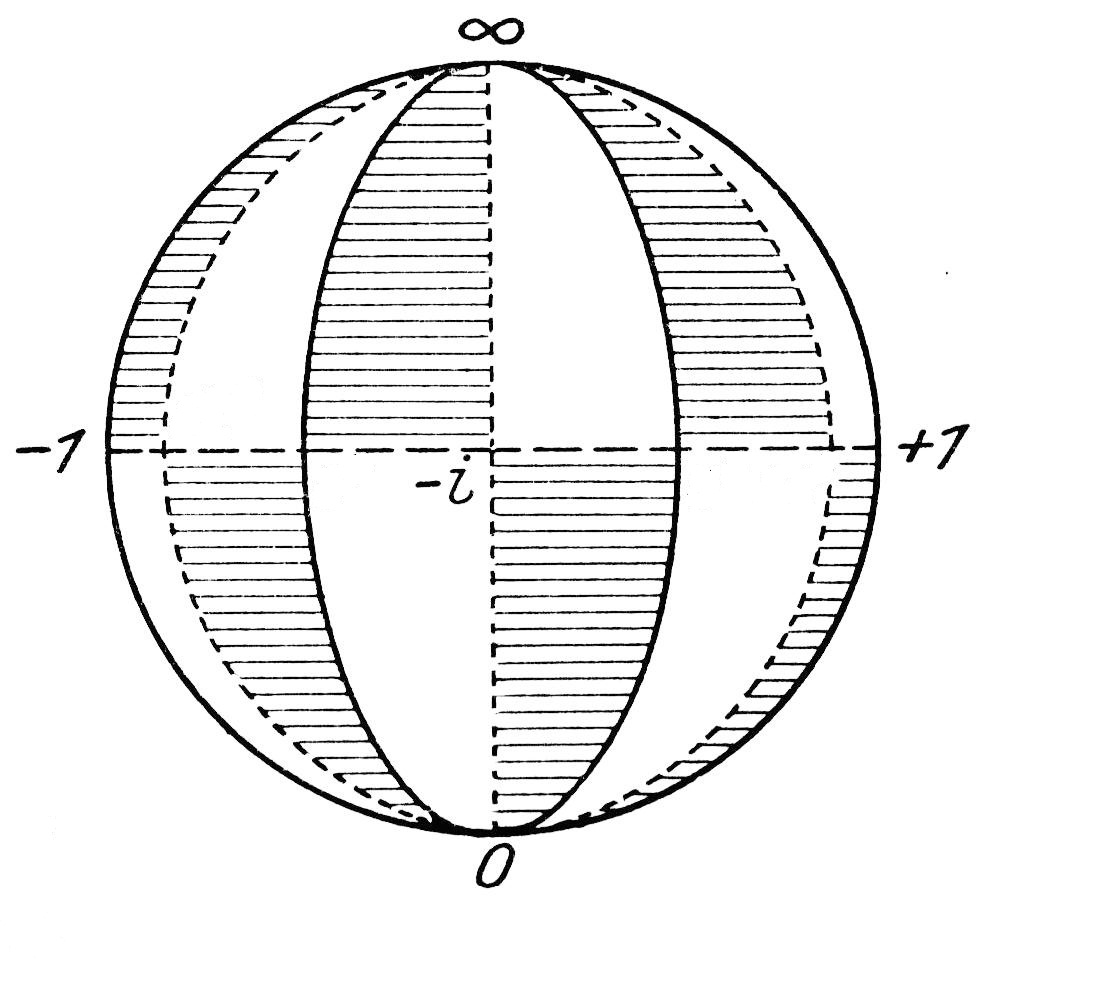}
		\caption{The $w$-sphere for the dihedron where the spherical triangles consist of hatched and unhatched parts with bases on the equator. The critical points, $-1$, $0$, $+1$, $\infty$, correspond to the branch points, $-1$, $+1$, and $\infty$ on the $z$-sphere.}
	\label{fig:dihedron}
\end{figure}

The spherical triangles are conformally represented in the $z$-half plane by the angular points, $-1,+1,\infty$. The plane is then projected onto a sphere stereographically~\cite{Forsyth}. The triangle on the sphere is  bounded by arcs of great circles that  cut  the equator orthogonally.  They form a double pyramid with the summits at the poles, each triangle having a vertex angle $(\ell-\alpha)\pi$, and  two right angles at the base on the equator, as shown in Fig.~\ref{fig:dihedron}.

At the pole, $\ell=\alpha$, the conformal mapping \eqref{eq:w} becomes
\begin{align*}
w&=\half\ln\left(\frac{z-\surd(z^2-1)}{z+\surd(z^2-1)}\right)\\
&=\ln\left[z-\surd(z^2-1)\right]\\
&=-\cosh^{-1}z.\end{align*}
Since $z=\cosh(-w)=\cosh w=\cosh(u+iv)$,
\[x=\cosh u\cos v\qquad\mbox{and}\qquad y=\sinh u\sin v,\]
the upper half of the $z$-plane is conformally mapped onto the interior of a strip in the $w$-plane bounded by the line $u=0$ on the right, and the lines $v=0$ on the bottom and $v=\pi$ on the top.

The partial wave amplitude,  \eqref{eq:orth}, then reduces to $(-1)^{\ell}/(2\ell+1)$ at the pole, so that the contribution from one Regge pole at $\ell=\alpha$ to the partial wave projection, as $k^2\rightarrow0$, is
\[A_{\ell}\sim(-1)^{(\ell+1)}\frac{4k^{2\ell}\ln(m/k)}{\pi(2\ell+1)}.\]

Inverting the conformal map \eqref{eq:w} gives the dihedral equation~\cite{Klein},
\begin{align}
z&=\half\left(w^{n}+\frac{1}{w^{n}}\right)\nonumber\\
&=\cosh(n\ln w),\label{eq:cosh}
\end{align}
where~\footnote{The most interesting cases for angular momenta are $n=2$, and $n\rightarrow\infty$, although we can choose some multiple $s$ of $(\ell-\alpha)^{-1}$, which can be made arbitrarily close to an integer whose sole effect is to change the number of iterates on the elliptic element, $g\rightarrow g^s$.} 
\begin{equation}
n=(\ell-\alpha)^{-1}.\label{eq:n}
\end{equation}   
Introducing homogeneous coordinates \eqref{eq:cosh} can be written as
\begin{equation}
\frac{z_1}{z_2}=\frac{w_1^{2n}+w_2^{2n}}{2w_1^nw_2^n}\ge1,\label{eq:ag}
\end{equation}
which is the arithmetic-geometric mean inequality.
The vanishing of the functional determinant of the numerator and denominator of \eqref{eq:ag}~\cite{Klein},
\[\left|\begin{matrix}2nw_1^{2n-1} & 2nw_2^{2n-1}\\ 2nw_1^{n-1}w_2^{n} & 2nw_1^{n}w_2^{n-1}\end{matrix}\right|=4n^2w_1^{n-1}w_2^{n-2}\left(w_1^{2n}-w_2^{2n}\right)=0,\]
indicates that there are $(n-1)$ zeros at $w_1=0$ and $w_2=0$, and another $2n$ zeros at
\[\left(\frac{w_1}{w_2}\right)^n=\pm1.\]
Therefore, we can write the ratio of homogeneous coordinates as either,
\begin{equation}\frac{w_1}{w_2}=g^{r},\qquad\mbox{or}\qquad \frac{w_1}{w_2}=g^{\prime}g^{r}\qquad (r=0,1,\ldots, n-1),\label{eq:rot}
\end{equation}
where the first term
\[g=e^{2i\pi/n},\qquad \qquad g^{\prime}=e^{i\pi/n},\]
is the $n$th root of unity. 

Since the values of the homogenous coordinates have absolute value of unity, they lie on the equator, and \eqref{eq:rot} correspond to rotations of the $w$-sphere either through angles $2\pi r/n$, or $3\pi r/n$. According to \eqref{eq:cosh}, the critical points $w=0$ and $w=\infty$ of the $w$-sphere correspond to $z=\infty$ of the $z$-sphere. In addition, the points $w=g^r$ correspond to the point $z=1$ on the $z$-sphere, while those of $w=g^{\prime}g^r$ all correspond to $z=-1$ on the same sphere. Hence, there will be three branch points, $-1,+1,\infty$ on the $z$-sphere---the same three singular points of the Legendre function of the second kind.

\section{Complex Angular Momentum}
For $\alpha$ close to $\ell$ we can perform a Taylor series expansion in terms of the energy difference $\Delta E=(E-E_{\ell})$, and retain only linear terms~\cite{Frautschi},
\begin{equation}\alpha\simeq\ell+\alpha^{\prime}_R\Delta E+i\alpha_I,\label{eq:Regge}
\end{equation}
where $\alpha_R=\mbox{Re}\;\alpha$, $\mbox{Im}\;\alpha_I=\alpha$, and the prime denotes differentiation with respect to the energy, evaluated at the energy $E_{\ell}$. It is clear from the dihedral equation, \eqref{eq:cosh}, that $\ell>\alpha$ when both are real. This also follows from the condition that the fixed energy dispersion relation,
\begin{equation}
A_{\ell}(z_0)=\frac{1}{\pi}\int_{z_{0}}^{\infty}Q_{\ell}(z^{\prime})A_{\ell}(z^{\prime})dz^{\prime}, \label{eq:A}
\end{equation}
converges. For large $z$, $A_{\ell}(z)\sim z^{\alpha}$~\cite[p. 148]{Frautschi}, while $Q_{\ell}(z)\sim z^{-(\ell+1)}$ for \emph{all\/} values of $\ell$, and not like $P_{\ell}(z)$ which requires $\mbox{Re}\;\ell\le\half$, so that \eqref{eq:A} will converge only when $\mbox{Re}\;\ell>\mbox{Re}\;\alpha$.

Introducing \eqref{eq:Regge} into \eqref{eq:cosh}, together with $w=re^{i\varphi}$, give
\begin{equation}
z=\cosh\left(\frac{\Delta E\ln r+\half\Gamma\varphi+i\left(\varphi\Delta E-\half\Gamma\ln r\right)}{ \alpha_R^{\prime}\left(\Delta E^2+\fourth\Gamma^2\right)}\right). \label{eq:z}
\end{equation}
The resonance width, 
\[\half\Gamma=\frac{\alpha_I}{\alpha^{\prime}_R},\]
vanishes below threshold, but $\alpha^{\prime}_R$ is always positive definite. Introducing $z=x+iy$ into \eqref{eq:z} it becomes clear that only below threshold do we have a conformal mapping~\cite{Bieberbach} of circles, $r=\mbox{const}$., in the $w$-plane, onto ellipses, 
\begin{equation}
\frac{x^2}{\cosh^2(\ln r/\alpha_R^{\prime}\Delta E)}+\frac{y^2}{\sinh^2(\ln r/\alpha_R^{\prime}\Delta E)}=1, \label{eq:ellipse}
\end{equation}
in the $z$-plane, whose semi-major axis and eccentricity are  $\cosh(\ln r/\alpha_R^{\prime}\Delta E)$ and $\mbox{sech}\;(\ln r/\alpha_R^{\prime}\Delta E)$, respectively. In other words,
\begin{equation}
z=\cosh\left(\frac{\ln r+i\varphi}{\alpha_R^{\prime}\Delta E}\right),\label{eq:Lehmann}
\end{equation}
represents an ellipse with semi-major axis  $\cosh(\ln r/\alpha_R^{\prime}\Delta E)$, and eccentric angle, $\varphi/\alpha_R^{\prime}\Delta E$. In high energy physics, \eqref{eq:Lehmann} would be referred to as a Lehmann ellipse~\cite{Lehmann}. 

Moreover, the straight lines $\varphi=\mbox{const}$., in the $w$-plane, are mapped onto hyperbolas,
\begin{equation}
\frac{x^2}{\cos^2(\varphi/\alpha_R^{\prime}\Delta E)}-\frac{y^2}{\sin^2(\varphi/\alpha_R^{\prime}\Delta E)}=1, \label{eq:hyperbola}
\end{equation}
in the $z$-plane, with semi-axes $|\cos(\varphi/\alpha_R^{\prime}\Delta E)|$ and $|\sin(\varphi/\alpha_R^{\prime}\Delta E)|$. The foci of the hyperbolas are the same as those of the ellipses \eqref{eq:ellipse}, viz., $z=\pm1$. The families of ellipses, \eqref{eq:ellipse} and hyperbolas, \eqref{eq:hyperbola}, that constitute a system of confocal conics is destroyed above threshold  by a mixing of $\ln r$ and $\varphi$ in the real and imaginary parts of the argument in \eqref{eq:z} due to the presence of a finite resonance width, $\Gamma$.

\section{Discontinuous Elementary Groups}
\begin{figure}[ht]
	\centering
		\includegraphics[width=0.3\textwidth]{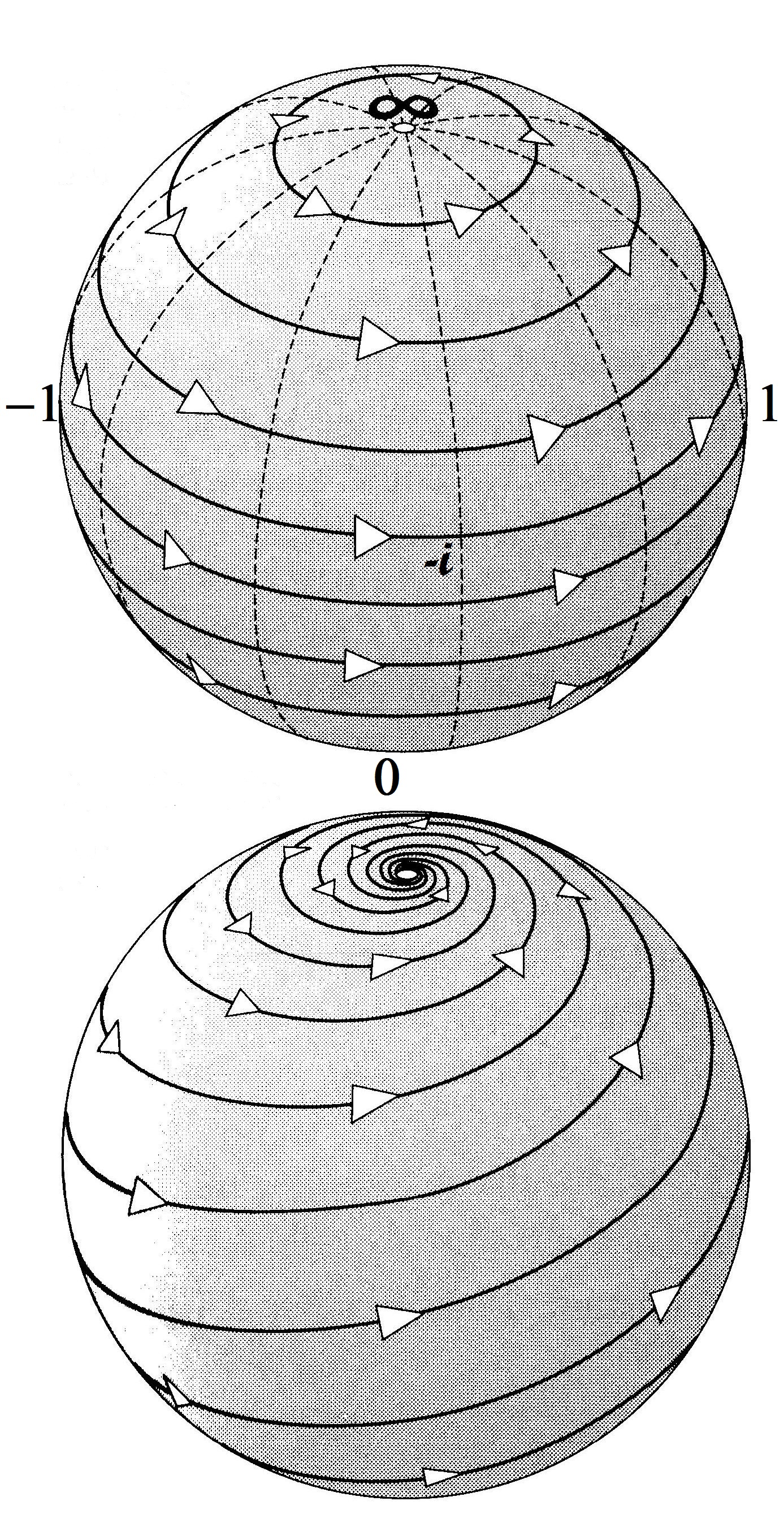}
	\caption{Elliptic (above) and loxodromic (below) motion.}
	\label{fig:loxo}
\end{figure}

Below threshold, the resonance width, $\Gamma$, vanishes identically, and the group of rigid motions are pure rotations,
\begin{equation} \mathsf{V}=\begin{pmatrix}\cos(\alpha_R^{\prime}\Delta E\pi) & i\sin(\alpha_R^{\prime}\Delta E\pi)\\ i\sin(\alpha_R^{\prime}\Delta E\pi) &\cos(\alpha_R^{\prime}\Delta E\pi)\end{pmatrix}.\label{eq:V}
\end{equation}
The rotation matrix \eqref{eq:V} ensures that the fixed points are at $\pm1$, instead of the usual rotation matrix which has indifferent fixed points, $\pm i$. The motion on a sphere is shown in the upper sphere in Fig.~\ref{fig:loxo}. We are in the physical region of bound states where $\cos\vartheta\in(-1,1)$.

Dynamical resonances for decaying unstable particles give a finite width above threshold. This introduces deformations which are hyperbolic elements,
\begin{equation}\mathsf{U}=\begin{pmatrix}\cosh\left(\half\alpha^{\prime}_R\Gamma\pi\right) & \sinh\left(\half\alpha^{\prime}_R\Gamma\pi\right)\\ \sinh\left(\half\alpha^{\prime}_R\Gamma\pi\right) & \cosh\left(\half\alpha^{\prime}_R\Gamma\pi\right) \end{pmatrix}.\label{eq:U}
 \end{equation}
 The fixed points of \eqref{eq:U} are the same as those of \eqref{eq:V}, and, consequently the trace of their commutator is $\mbox{tr}[U,V]=2$. The combination of stretching and rotation give rise to loxodromic elements,
\begin{equation}\mathsf{T}=\mathsf{U}\mathsf{V}=\begin{pmatrix}\cosh\alpha^{\prime}_R\left(\half\Gamma+i\Delta E\right)\pi& \sinh\alpha^{\prime}_R\left(\half\Gamma+i\Delta E\right)\pi\\ \sinh\alpha^{\prime}_R\left(\half\Gamma+i\Delta E\right)\pi & \cosh\alpha^{\prime}_R\left(\half\Gamma+i\Delta E\right)\pi\end{pmatrix},\label{eq:UV}
 \end{equation}
which can be thought of as a rotation through a complex angle, $\alpha_R^{\prime}(\Delta E-i\half\Gamma)\pi$.

 The map,
 \[z\mapsto\frac{z+1}{z-1},\]
 moves $-1$ to $0$ and $+1$ to $\infty$. Mathematically, this is mere conjugation, but, physically, it represents the transformation of real (bound states) into complex (dynamical resonances) angular momenta. The loxodromic motion is shown in the lower part of Fig.~\ref{fig:loxo}, which is described by the scaling map,
\begin{equation}
z\mapsto az, \label{eq:loxodromic}
\end{equation}
where the modulus of the multiplier, $a=e^{\alpha_R^{\prime}(\Gamma+i2\Delta E)}$, is greater than $1$.  
The fixed point at $\infty$ (or $+1$) is an attracting fixed point, or sink, while the fixed point at $0$ (or $-1$) is a repelling fixed point, or source. Therefore, points will spiral out of the source, $0$, and spiral into the sink, $\infty$, as shown in lower figure in Fig.~\ref{fig:loxo}.  Below threshold, $\Gamma=0$, and the fixed points are indifferent, giving rise to the pattern shown in the upper figure in Fig.~\ref{fig:loxo}.

For $\ell$ a positive integer, the contour $\mathcal{C}_1$ in the definition of the Legendre function of the first kind~\cite{Courant},
\begin{equation}
P_{\ell}(z)=\frac{1}{2\pi i}\int_{\mathcal{C}_{1}}\frac{1}{2^{\ell}}\frac{(\zeta^2-1)^{\ell}}{(\zeta-z)^{\ell+1}}d\zeta,\label{eq:P}
\end{equation}
 is a closed curve, oriented in the positive direction, that encloses both points $z$. The integrand is real when $z$ is on the real axis between $-1$ and $+1$. This is the physical region. For non-integer $\ell$, \eqref{eq:P} is a hypergeometric function with a branch cut going from $-1$ to $-\infty$ in order to keep it single-valued. 

In contrast, Legendre function of the second kind,
\begin{equation}
Q_{\ell}(z)=\frac{1}{4i\sin\ell\pi}\int_{\mathcal{C}_{2}}\frac{1}{2^{\ell}}
\frac{\zeta^2-1)^{\ell}}{(z-\zeta)^{\ell+1)}}d\zeta, \label{eq:Q}
\end{equation}
where $\mathcal{C}_2$ is a figure $8$ encircling the points $-1$ and $+1$. The Legendre function of the second kind, \eqref{eq:Q}, is regular and single-valued in the $z$-plane which has been cut along the real axis from $1$ to $-\infty$. According to \eqref{eq:Q}, $\ell$ cannot be an integer, and it has logarithmic singularities at $z=1$ and $z=-1$ because the path must cut the lines joining $z$ to $+1$ and $z=-1$.  

The isometric circles of $\mathsf{U}$ and $\mathsf{U}^{-1}$ are \[\mathsf{I}:|z+e^{\half\alpha_R^{\prime}\Gamma\pi}|=0\qquad \mbox{and}\qquad \mathsf{I}^{\prime}:|z+e^{-\half\alpha^{\prime}_R\Gamma\pi}|=0,\] respectively, which are exterior to one another. Isometric circles define a complete locus of points in the neighborhood of which lengths and areas remain invariant under substitutions of the form $\mathsf{T}$~\cite{Ford}. It can be shown that the fundamental region for the group of motions generated by $\mathsf{T}$ consists of that part of the plane exterior to $\mathsf{I}$ and $\mathsf{I}^{\prime}$~\cite[p. 54]{Ford}. A point belongs to the fundamental region if a circle can be drawn about the point as its center that does not contain an interior of the isometric circles.

For both real and complex $(\ell-\alpha)$, the groups which are formed are discrete and elementary. By elementary it is meant that there exist finite orbits in $\mathbb{R}^3$~\cite{Beardon}. In the case that of real angular momenta, they constitute the group of finite, cyclic rotations. More specifically, if $s$ is  number of vertices of the polygon, where each vertex has $n_j$ elements, then the relation to the size of the group, $|G|$, and the number of elements of the vertices is~\cite[p. 85]{Beardon}
\begin{equation}
2\left(1-\frac{1}{|G|}\right)=\sum_{j=1}^s\left(1-\frac{1}{n_{j}}\right).\label{eq:G}
\end{equation}
Moreover, since $n_j\ge2$ it also follows that the right-hand side of \eqref{eq:G} is bounded by
\[\half s\le\sum_{j=1}^s\left(1-\frac{1}{n_{j}}\right)<s.\]

Now, for $s=2$ we have a crescent for by two non concentric intersecting circles with two vertices, and \eqref{eq:G} reduces to
\[\frac{1}{|G|}=\half\left(\frac{1}{n_{1}}+\frac{1}{n_{2}}\right),\]
which identifies $|G|$ as the harmonic mean. It can be satisfied by $|G|=n_1=n_2$. By conjugation, the fixed points, or vertices, can be brought to $0$ and $\infty$. Then $G$ is a finite, cyclic group of rotations in $\mathbb{R}^2$~\cite{Beardon}.

The case $s=3$ applies to a dihedron, for which \eqref{eq:G} becomes
\begin{equation}
\sum_{j=1}^s\frac{1}{n_{j}}=1+\frac{2}{|G|}.\label{eq:G-bis}
\end{equation}
Suppose the number of elements at the vertices are ordered as $n_1\le n_2\le n_3$. The choice $n_1=3$ is incompatible with \eqref{eq:G-bis} since the sum on the left will be inferior to $1$. We, therefore, choose $n_1=2$, and \eqref{eq:G-bis} reduces to
\[\frac{1}{n_2}+\frac{1}{n_3}=\frac12+\frac{2}{|G|}.\]
Since $n_2\ge4$ will lead to a contradiction, $n_2$ can be either $2$ or $3$. In the former case $n_3$ is free to take on any value from $2$ to $\infty$ so there will be $|G|=2n$ sides of the various triangles in the division of the sphere, as shown in Fig.~\ref{fig:dihedron}. 

Now another elementary group has $n=2$, and has every element of the group leaving $0$ and $\infty$ invariant~\cite[p. 84]{Beardon}. The group cannot have the Poincar\'e metric, $dz/(1-|z|^2)$,  because loxdromic motion does not preserve the unit disc, $\mathbb{D}^2$. Rather, what is required is Poincar\'e's extension which adds on an additional dimension so that points on the  plane $\mathbb{R}^2$ are stereographically projected  onto the sphere $\mathbb{S}^2$. Rotations, like loxodromic transformations, also do not have an invariant \emph{real\/}  disc, $\mathbb{D}^2$, so that both elliptic and loxodromic generators live on the sphere, $\mathbb{S}^2$. 

What appears mathematically as conjugation, by transferring the fixed points $-1$ and $+1$ to $0$ and $\infty$, is physically quite different for it takes the system from real to complex angular momenta. In other words, merely by conjugation, the vertices, $-1,+1$, transform the physical region $z\in(-1,+1)$, or $k^2<0$ in \eqref{eq:cos}, into $0$ and $\infty$, of the unphysical region, $z\notin(-1,+1)$, or $k^2>0$. The two types of transformations will give a different tessellation of hyperbolic space, and so can be studied in much the same way that Poincar\'e characterized his \emph{kleinian\/} groups, which are much less well-known than their \emph{fuchsian\/} counterparts~\cite[Ch. 8]{Beardon}.

\section{Strengths and Weaknesses of the Analysis}
Different definitions of the Legendre function may be adopted according to whether $z$ is considered to be an unrestricted complex variable, or a real variable confined to the open interval $z\in(-1,1)$. However, there is no relation between complex values of the degree of a Legendre function and the unphysical region of $z\notin(-1,1)$.

According to \eqref{eq:cos}, bound states, $k^2<0$, lie in the physical region. But, what is missing is a conformal mapping, 
\begin{equation}
\ell=\alpha(k^2),\label{eq:Regge-bis}
\end{equation} 
between the $\ell$-plane and the $\kappa$-plane. If we go back to the Watson-Sommerfeld representation, \eqref{eq:WS}, we can write it, in the vicinity of $k_{\ell}$, as~\cite[p. 46]{Omnes}
\begin{equation}
\frac{\beta(k)P_{\alpha(k^2)}(-\cos\vartheta)}{\sin\pi\alpha(k^2)}=\frac{(\beta(k_{\ell})P_{\ell}(-\cos\vartheta)}{\pi\alpha^{\prime}(k_{\ell}^2)\left(k^2-k_{\ell}^2\right)\cos\pi\ell}+\mbox{regular function}, \label{eq:WS-bis}
\end{equation}
where $\alpha$ is considered a real function of $k^2$, the energy. Thus, if we consider the conformal mapping
\[w=\sin\pi\alpha(k^2),\]
then a semi-infinite strip of width $1$ in the $\alpha$-plane is mapped onto the upper half $w$-plane where the boundaries of the strip are homologues of the points $-1$ and $+1$ on the real axis. The poles of \eqref{eq:WS-bis} will then be at $\ell=\pm\half$.

It is assumed that for $k^2<0$, the Regge trajectories \eqref{eq:Regge-bis} are real, but once $k^2>0$, the Regge trajectories become complex. This being a bound state would lead to the conclusion that Regge trajectories are \lq\lq the desired interpolating functions between bound states\rq\rq~\cite{Omnes-bis}.

For $k^2>0$, $\alpha(k^2)$ is complex, and this describes resonances. It is argued~\cite{Omnes-bis} that $\mbox{Re}\;\alpha$ can coincide with integer $\ell_0$ for $k^2>0$, while $\mbox{Im}\;\alpha(k^2)$ remains small. Then appealing to \lq continuity\rq\ $k^2$ is considered complex where $\alpha$ becomes exactly equal to $\ell_0$. Consequently, we can take $\alpha$ physical with unphysical $k^2$, which is a resonance, or unphysical $\alpha$ with physical $k^2$, which would correspond to a metastable state known as a \lq shadow\rq\ state~\cite[p. 47]{Omnes}. However,  we are not at liberty to consider $k^2$ complex as we originally considered $k$ complex.

Rather,  discrete, elementary groups can be employed to distinguish between bound and resonance states. Only for maps of the form
\[z\mapsto az^s, \qquad |a|\neq1, \quad s^2=1,\]
will the angular momentum become complex, and so transform the indifferent fixed points of Type 1 group into a Type 3 group whose orbits leave invariant the  source-sink fixed points at $0$ and $\infty$~\cite[p. 84]{Beardon}. What is still missing is the form of the Regge trajectory \eqref{eq:Regge-bis} which would connect $\mbox{Re}\;\alpha$ with $k^2<0$, and so lead to physical values in \eqref{eq:cos}, and complex values of $\alpha$ with $k^2>0$ which result in unphysical values of \eqref{eq:cos}.


\begin{thebibliography}{99}
\bibitem{Khuri}N. N. Khuri, \emph{Phys. Rev\/}. {\bf 130} (1963) 429.
\bibitem{Jones}E. Jones, Lawrence Radiation Laboratory Report UCRL-10700 (1963).
\bibitem{Omnes}R. Omn\`es \& M. Froissart, \emph{Mandelstam Theory and Regge Poles\/} (Benjamin, New York, 1963), \S 4.2.
\bibitem{Poincare}H. Poincar\'e, \emph{Rendiconti Circolo Mat. Palermo\/} {\bf 29} (1910) 169.
\bibitem{Barut}A. O. Barut \& D. E. Zwanziger, \emph{Phys. Rev\/}. {\bf 127} (1962) 974.
\bibitem{Newton}R. G. Newton, \emph{J. Math. Phys\/}. {\bf 3} (1962) 867.
\bibitem{Erdelyi}A. Erdelyi, W. Magnus, F. Oberhettinger, \& F. G. Tricomi (eds.) \emph{Higher Transcendental Functions\/}, vol. 1 (McGraw-Hill, New York, 1953), p. 162.
\bibitem{Forsyth}A. R. Forsyth, \emph{Theory of Functions of a Complex Variable\/}, vol. II, 3rd ed. (Cambridge U. P., Cambridge, 1918), p. 692.
\bibitem{Klein}F. Klein, \emph{Elementary Mathematics from an Advanced Standpoint\/} (Macmillan, New York, 1932), pp. 115-117.
\bibitem{Frautschi}S. Frautschi, \emph{Regge Poles and S-Matrix Theory\/} (Benjamin, New York, 1963), p. 114.
\bibitem{Bieberbach}L. Bieberbach, \emph{Conformal Mapping\/} (Chelsea, New York, 953), \S 12.
\bibitem{Lehmann}H. Lehmann, \emph{Nuovo Cimento\/} {\bf X} (1958) 579.
\bibitem{Courant}See, for example, D. Hilbert and R. Courant, \emph{Methods of Mathematical Physics\/}, vol. I (Interscience, New York, 1953), pp. 501-508.
\bibitem{Ford}L. R. Ford, \emph{Automorphic Functions\/}, 2nd ed. (Chelsea, New York, 1929), p. 25.
\bibitem{Beardon}A. F. Beardon, \emph{The Geometry of Discrete Groups\/} (Springer, New York, 1983), p. 82.
\bibitem{Omnes-bis}R. Omn\`es, \emph{Introduction to Particle Physics\/} (Wiley-Interscience, New York, 1971), p. 347.
\end{thebibliography}
\end{document}